\documentclass[sigconf]{acmart}

\usepackage{adjustbox}
\usepackage{hhline}
\usepackage{multirow}
\usepackage{url}
\usepackage{balance}
\setcopyright{acmlicensed}
\acmPrice{15.00}
\acmDOI{10.1145/3121245.3121249}
\acmYear{2017}
\copyrightyear{2017}
\acmISBN{978-1-4503-5155-3/17/09}
\acmConference[A-TEST'17]{8th International Workshop on Automated Software Testing}{September 4-5, 2017}{Paderborn, Germany}

\begin{document}
\title{Dynamic Mutant Subsumption Analysis using LittleDarwin}

\author{Ali Parsai}
\orcid{0000-0001-8525-8198}
\affiliation{%
  \institution{University of Antwerp}
  \streetaddress{Middelheimlaan 1}
  \city{Antwerp} 
  \country{Belgium}
  \postcode{2020}
}
\email{ali.parsai@uantwerpen.be}

\author{Serge Demeyer}
\orcid{0000-0002-4463-2945}
\affiliation{%
    \institution{University of Antwerp}
    \streetaddress{Middelheimlaan 1}
    \city{Antwerp} 
      \country{Belgium}
      \postcode{2020}
}
\email{serge.demeyer@uantwerpen.be}

\begin{abstract}
Many academic studies in the field of software testing rely on mutation testing to use as their comparison criteria. However, recent studies have shown that redundant mutants have a significant effect on the accuracy of their results. One solution to this problem is to use mutant subsumption to detect redundant mutants. Therefore, in order to facilitate research in this field, a mutation testing tool that is capable of detecting redundant mutants is needed. In this paper, we describe how we improved our tool, LittleDarwin, to fulfill this requirement. %
\end{abstract}

\ccsdesc[500]{Software and its engineering~Software testing and debugging}
\setcopyright{acmlicensed}
\acmPrice{15.00}
\acmDOI{10.1145/3121245.3121249}
\acmYear{2017}
\copyrightyear{2017}
\acmISBN{978-1-4503-5155-3/17/09}
\acmConference[A-TEST'17]{8th International Workshop on Automated Software Testing}{September 4-5, 2017}{Paderborn, Germany}

\keywords{Software Testing, Mutation Testing, Mutant Subsumption, Dynamic Mutant Subsumption}

\maketitle

\section{Introduction}
\label{S:Intro}
Many academic studies on fault detection need to assess the quality of their technique using seeded faults. One of the widely-used systematic ways to seed simulated faults into the programs is mutation testing~\cite{DeMillo1978}.  
Mutation testing is the process of injecting faults into software (i.e. creating a mutant), and counting the number of these faults that make at least one test fail (i.e. kill the mutant). The process of creating a mutant consists of applying a predefined transformation on the code (i.e. mutation operator) that converts a version of the code under test into a faulty version. It has been shown that mutation testing is an appropriate method to simulate real faults and perform comparative analysis on testing techniques~\cite{Andrews2005,Andrews2006,Just2014}.

There has been many studies to optimize the process of mutation testing by following the maxim \textit{\{do faster, do smarter, do fewer\}}~\cite{Offutt2001}. In particular, \textit{do fewer} aims to reduce the number of produced mutants. There are several techniques that implement this logic (e.g. selective mutation~\cite{Mathur1991,Offutt1993,Offutt1996}, and mutant sampling~\cite{Wong1995,Zhang2010,Zhang2013,Parsai2016}). However, only recently the academics began to investigate the threats to validity the redundant mutants introduce in software testing experiments~\cite{Papadakis2016}. Papadakis et al. demonstrate that the existence of redundant mutants introduces a significant threat by ``artificially inflating the apparent ability of a test technique to detect faults''~\cite{Papadakis2016}. 

One of the recent solutions to alleviate this problem is to use mutant subsumption~\cite{Ammann2014}. Mutant \emph{A} \textit{truly} subsumes mutant \emph{B} if and only if all inputs that kill \emph{A} also kill \emph{B}. This means that mutant \emph{B} is redundant, since killing \emph{A} is sufficient to know that \emph{B} is also killed. It is possible to provide a more accurate analysis of a testing experiment by determining and discarding the redundant mutants. However, it is often impossible to check mutants for every possible input to the program in practice. Therefore, as a compromise, dynamic mutant subsumption is used instead~\cite{Ammann2014}.  Mutant \emph{A} \textit{dynamically} subsumes mutant \emph{B} with regards to test set \emph{T} if and only if there exists at least one test that kills \emph{A}, and every test that kills \emph{A} also kills \emph{B}. 
Given the fact that mutant subsumption  only recently  has been at the center of attention, there are no mature tools that can perform dynamic mutant subsumption analysis on real-life Java programs. This, however, is necessary to facilitate further research on the topic. Therefore we aim to fill this void by developing such tool.

We used LittleDarwin\footnote{\url{https://littledarwin.parsai.net/}} mutation testing framework to implement the features needed to perform dynamic mutant subsumption analysis. LittleDarwin is an extensible and easy to deploy mutation testing tool for Java programs~\cite{Parsai2017}. LittleDarwin has  been used previously in several other studies~\cite{Parsai2016,Parsai2016M}, and it is shown to be capable of analyzing large and complicated Java software systems~\cite{Parsai2015T}.

The rest of the paper is organized as follows: 
In Section~\ref{S:Background}, background information about mutation testing is provided. 
In Section~\ref{S:SOTA}, the current state of the art is discussed.
In Section~\ref{S:DMSA}, we provide details on how LittleDarwin can help performing dynamic mutant subsumption analysis. 
Finally, we present our conclusions in Section~\ref{S:Conclusion}.

\section{Background}
\label{S:Background}
The idea of mutation testing was first mentioned by Lipton, and later developed by DeMillo, Lipton and Sayward~\cite{DeMillo1978}. The first implementation of a mutation testing tool was done by Timothy Budd in 1980~\cite{Budd1980}.
Mutation testing is performed as follows: First, a faulty version of the software is created by introducing faults into the system \textit{(Mutation)}. This is done by applying a known transformation \textit{(Mutation Operator)} on a certain part of the code.  After generating the faulty version of the software \textit{(Mutant)}, it is passed onto the test suite. If there is an error or failure during the execution of the test suite, the mutant is marked as killed \textit{(Killed Mutant)}. If all tests pass, it means that the test suite could not catch the fault, and the mutant has survived \textit{(Survived Mutant)}~\cite{Jia2011}.

If the output of a mutant for all possible input values is the same as the original program, it is called an \emph{equivalent mutant}. It is not possible to create a test case that passes for the original program and fails for an equivalent mutant, because the equivalent mutant is indistinguishable from  the original program. This makes the  creation of equivalent mutants undesirable, and leads to false positives during mutation testing.  In general, detection of equivalent mutants is undecidable due to the halting problem~\cite{Offutt1997}. Manual inspection of all mutants is the only way of filtering all equivalent mutants, which is impractical in real projects due to the amount of work it requires. Therefore, the common practice within today's state-of-the-art is to take precautions to generate as few equivalent mutants as possible, and accept equivalent mutants as a threat to validity (accepting a false positive is less costly than removing a true positive by mistake~\cite{Fawcett2006}). 

\begin{equation}
Mutation\ Coverage = \frac{Number\ of\ killed\ mutants}{Number\ of\ all\ non\mbox{-}equivalent\ mutants}
\label{coverageequation}
\end{equation}

Mutation testing allows software engineers to monitor the fault detection capability of a test suite by means of mutation coverage (see Equation~\ref{coverageequation})~\cite{Jia2011}.
A test suite is said to achieve \textit{full mutation test adequacy} whenever it can kill all the non-equivalent mutants, thus reaching a mutation coverage of 100\%. Such test suite is called a \textit{mutation-adequate test suite}. 

\section{State of the Art}
\label{S:SOTA}

\begin{figure}
    \centering	
    \fbox{
        \includegraphics[width=0.9\linewidth]{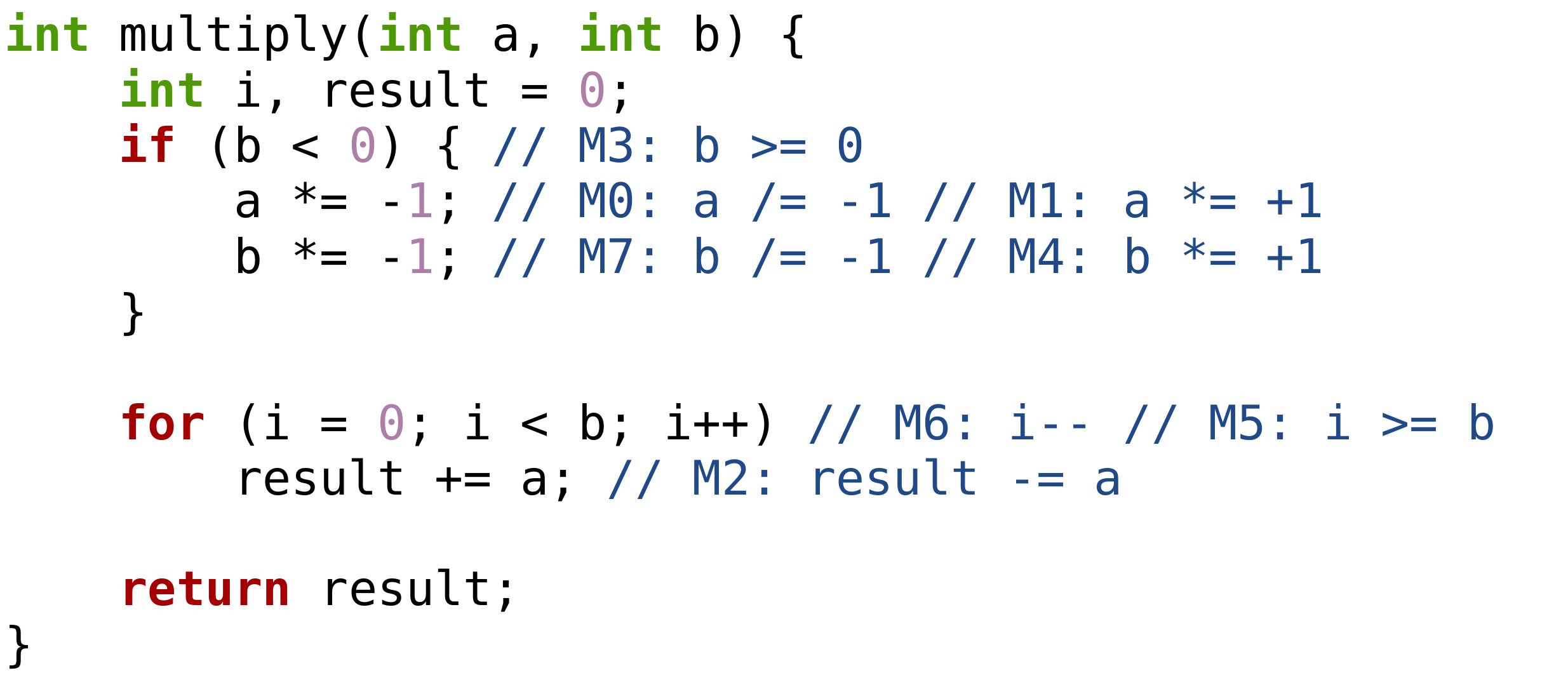}}
    \caption{An Example Mutated Method}
    \label{fig:dmsexample}
\end{figure}

Mutant subsumption is defined as the relationship between two non-equivalent mutants \texttt{A} and \texttt{B} in which \texttt{A} subsumes \texttt{B} if and only if all inputs that kill \texttt{A} is guaranteed to kill \texttt{B}~\cite{Kurtz2015}. The subsumption relationship for faults has been defined by Kuhn in 1999~\cite{Kuhn1999}, but its use for mutation testing has been popularized by Jia et al. for creating hard to kill higher-order mutants~\cite{Jia2008}. Later on, Ammann et al.  tackled the theoretical side of mutant subsumption~\cite{Ammann2014}. In their paper, Ammann et al. define \textit{dynamic} mutant subsumption, which redefines the relationship using test cases. Mutant \texttt{A} dynamically subsumes Mutant \texttt{B} if and only if (i) \texttt{A} is killed, and (ii) every test that kills \texttt{A} also kills \texttt{B}.
Kurtz et al.~\cite{Kurtz2015} use the notion of dynamic mutant subsumption graph (DMSG) to visualize the concept of dynamic mutant subsumption. Each node in a DMSG represents a set of all mutants that are mutually subsuming. Edges in a DMSG represent the dynamic subsumption relationship between the nodes. They introduce the concept of static mutant subsumption graph, which is a result of determining the subsumption relationship between mutants using static analysis techniques.

\begin{table}
    \centering
    \caption{Range of Input Values that Kill Mutants of the Example Mutated Method (left), DMSG for the Example Mutated Method (right)}
    \label{table:dmsexamplerange}
    \begin{minipage}{0.84\linewidth}
    \begin{tabular}{|c|c|c|}
        \hline \textbf{Mutants}    & \textbf{Range of $a$} & \textbf{Range of $b$} \\ 
        \hline \textbf{M0}   & $\varnothing$ & $\varnothing$ \\ 
        \hline \textbf{M1}   & $(-\infty, \infty) - \{0\}$ & $(-\infty, 0)$ \\ 
        \hline \textbf{M2}   & $(-\infty, \infty) - \{0\}$ & $(-\infty, \infty) - \{0\}$ \\ 
        \hline \textbf{M3}   & $(-\infty, \infty) - \{0\}$ & $(-\infty, \infty) - \{0\}$ \\ 
        \hline \textbf{M4}   & $(-\infty, \infty) - \{0\}$ & $(-\infty, 0)$  \\ 
        \hline \textbf{M5}   & $(-\infty, \infty) - \{0\}$ & $(-\infty, \infty)$ \\ 
        \hline \textbf{M6}   & $(-\infty, \infty) - \{0\}$ & $(-\infty, \infty) - \{0\}$ \\
        \hline \textbf{M7}   & $\varnothing$ & $\varnothing$ \\
        \hline
    \end{tabular} 
    \end{minipage}
    \begin{minipage}{0.15\linewidth}
    \includegraphics[width=\textwidth]{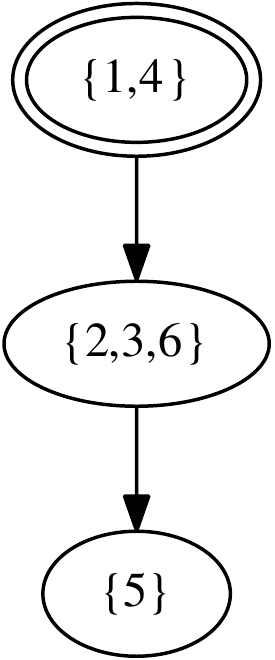}
    \end{minipage}
        
\end{table}

Figure~\ref{fig:dmsexample} shows a Java method and its set of mutants. This method takes $a$ and $b$ as input, and returns $a \times b$ as output. To do this, $a$ is added $b$ times. If $b$ is negative, both $a$ and $b$ are negated so that $b$ becomes positive. Table~\ref{table:dmsexamplerange} shows the range of input values that kills each mutant. As the table shows, M0 and M7 are equivalent mutants, since the change they introduce does not impact the program semantically. M1 and M4 are killed by the same range of inputs. The same holds true for  M2, M3, and M6. It can be seen that \{M1,M4\} truly subsume \{M2,M3,M6\}, since any input that kills M1 or M4, also kills M2, M3, and M6; however, the opposite does not hold. Also, \{M2,M3,M6\} truly subsume \{M5\} for the same reason. Using a test suite that includes a test case from each of the input ranges in Table~\ref{table:dmsexamplerange}, it is possible to draw the DMSG for this method.

The main purpose behind the use of mutant subsumption is to reliably detect redundant mutants, which create multiple threats to the validity of mutation testing~\cite{Papadakis2016}. This is often done by determining the dynamic subsumption relationship among a set of mutants, and keeping only those that are not subsumed by any other mutant. In our example, keeping only  M1 (or M4) suffices, since it subsumes all the other mutants. 

\begin{figure*}[!h]
    \centering
    \includegraphics[width=\linewidth]{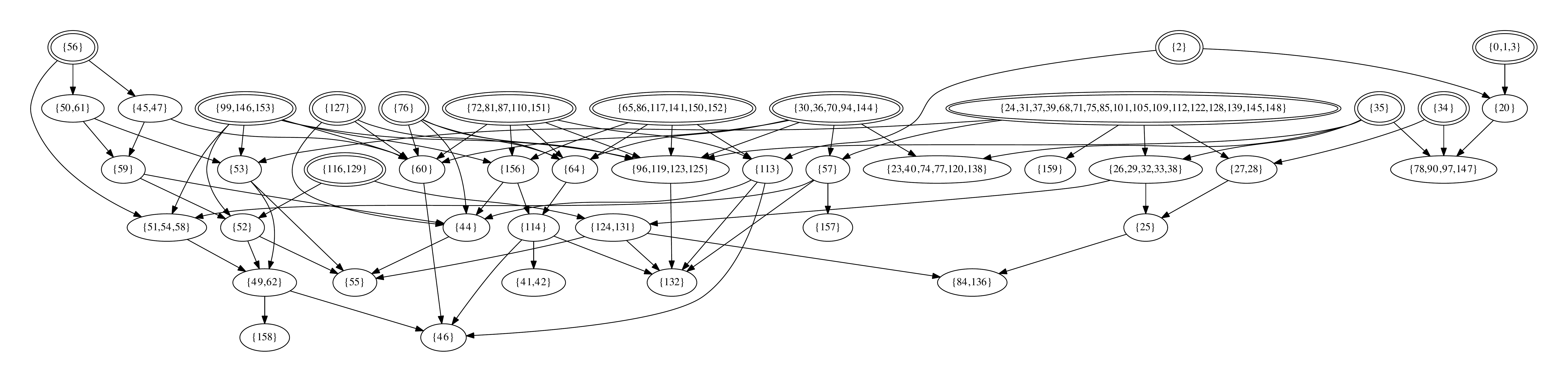}
    \caption{Dynamic Mutant Subsumption Graph for JTerminal}
    \label{fig:jterminal}
\end{figure*}

\section{Dynamic Mutant Subsumption Analysis with LittleDarwin}
\label{S:DMSA}

\begin{figure}
    \centering
    \includegraphics[width=0.7\linewidth]{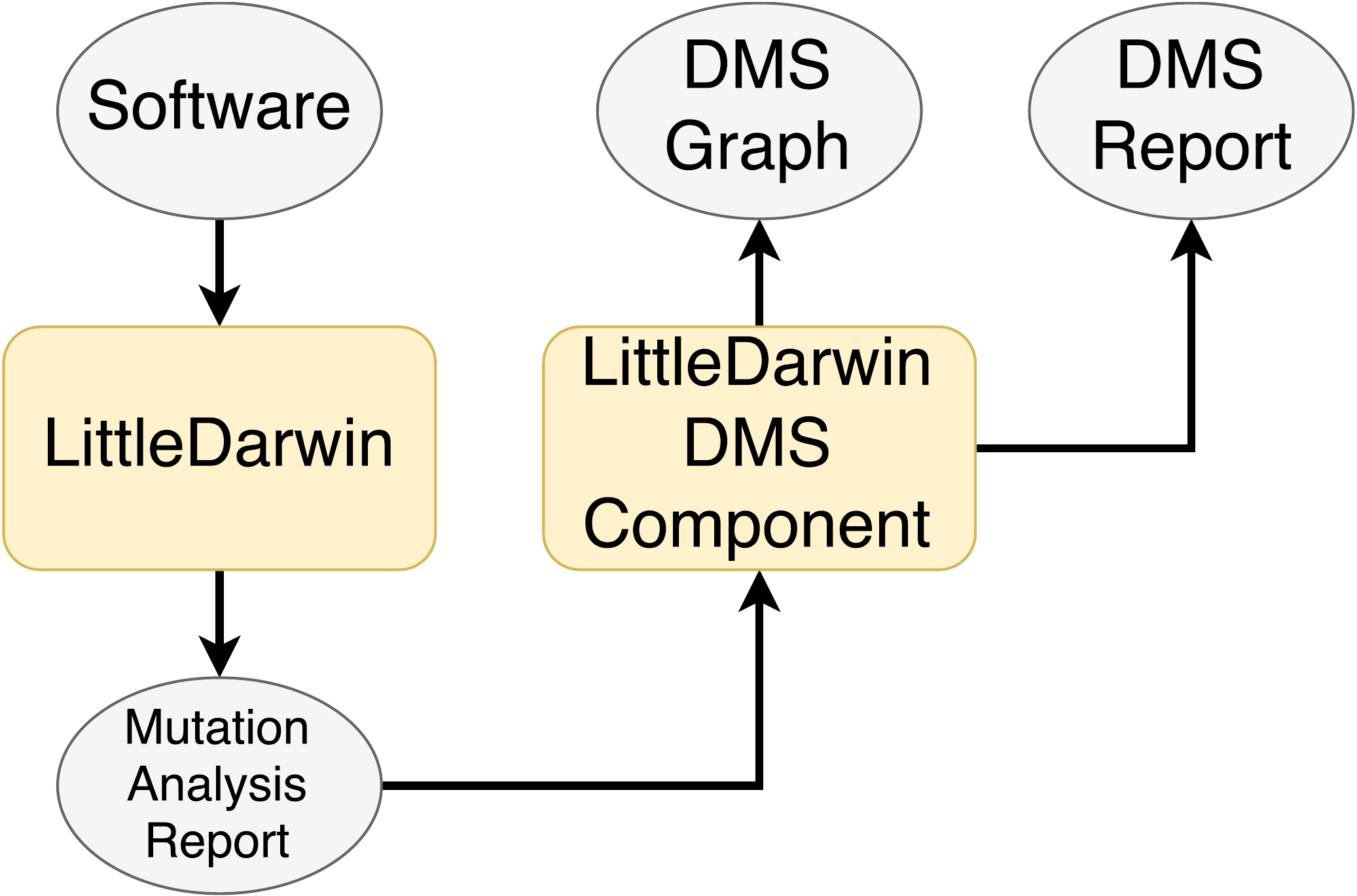}
    \caption{Dynamic Mutant Subsumption Component I/O}
    \label{fig:dms}
\end{figure}

Figure~\ref{fig:dms} shows the input and output of LittleDarwin's dynamic mutant subsumption (DMS) component. 
To facilitate dynamic mutant subsumption analysis in LittleDarwin, we retain all the output provided by the build system for each mutant. As a result, we  can parse this output and extract useful information, e.g. which test cases kill a particular mutant. 
LittleDarwin's DMS component can then use this information to determine dynamic subsumption relation between each mutant pair. This component then outputs the results in two different ways: (i) the dynamic mutant subsumption graph, to visualize the subsumption relation, and (ii) %
a detailed report is generated in CSV\footnote{Comma-separated Values} format that contains all the information processed by the DMS component. For each mutant, mutant ID, mutant path, source path, mutated line number, whether it is a subsuming mutant, number of failed tests, the mutants it subsumes, the mutants that it is subsumed by, and the mutants that are mutually subsuming with it are provided in this report.   Since LittleDarwin is a Java mutation testing framework, the application of the DMS component is also restricted to Java programs.

\begin{figure}
    \centering
    \fbox{
        \includegraphics[width=0.9\linewidth]{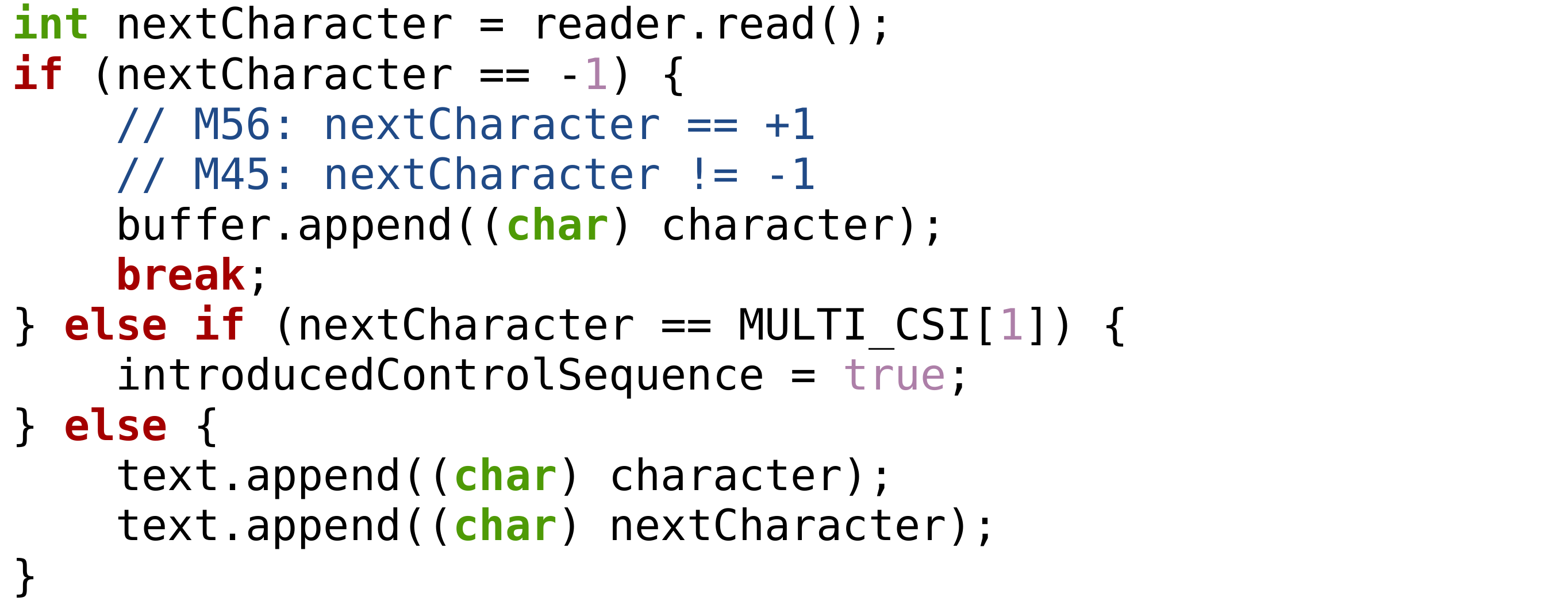}}
    \caption{Mutants 45 and 56 of JTerminal}
    \label{fig:code}
\end{figure}

\begin{figure}
    \centering
    \fbox{
        \includegraphics[width=0.9\linewidth]{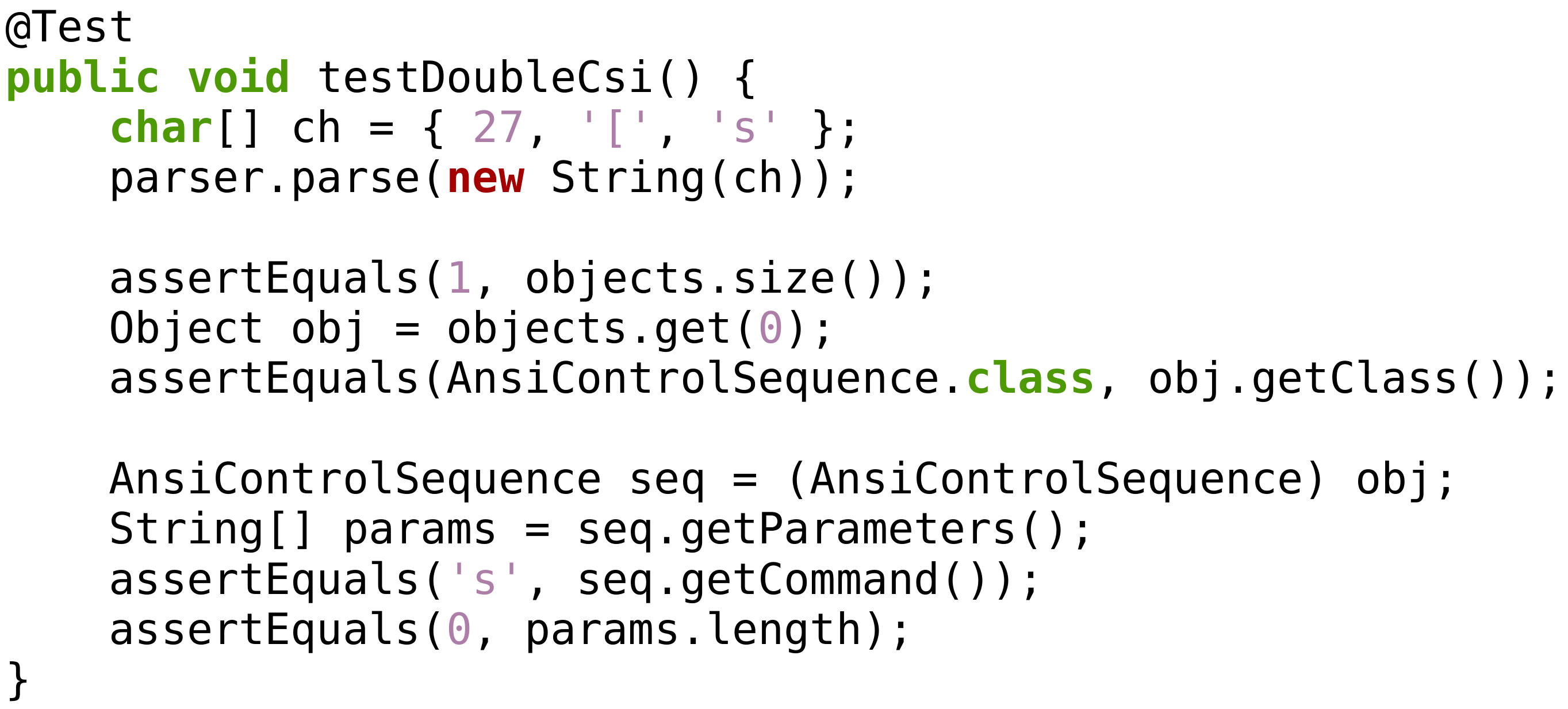}}
    \caption{The Test that Kills Mutant 45, but Not Mutant 56}
    \label{fig:testcode}
\end{figure}

To showcase the ability of LittleDarwin in performing dynamic mutant subsumption analysis, we use JTerminal\footnote{\url{https://www.grahamedgecombe.com/projects/jterminal}} as a subject.
We improved the test suite of JTerminal by automatically generating test cases using EvoSuite~\cite{SBST17_competition}. 
The information about characteristics of JTerminal is shown in Table~\ref{table:cases}. 
The DMSG for JTerminal is depicted in Figure~\ref{fig:jterminal}. In this figure, each number represents a single killed mutant, each node represents a group of mutants that are killed by exactly the same set of test cases, and each edge shows the dynamic subsumption relationship between each node where the node at the end is subsumed by the node at the start. The survived mutants are not shown in this figure. 
The double-circled nodes contain the subsuming mutant groups. In order to remove the redundant mutants, one only needs to keep one mutant from each subsuming mutant group and discard the rest. 

Take M45 and M56 as an example. According to the DMSG, M56 subsumes Mutant M45. Using the CSV report, we can locate the actual mutation of the source code (Figure~\ref{fig:code}). Both M45 and M56 belong to method \texttt{parse}     of class \texttt{AnsiControlSequenceParser}, and mutate the same statement on line 99. %
M45 acts as a negation of the conditional statement. This means that any input character (except -1) that used to trigger "else if" and "else", now trigger this branch. Since this branch contains a "break" statement, it avoids the rest of the iteration of the loop to be executed. If the input is -1, the "else" branch would be executed, which wrongfully appends -1 to "text".
M56, however, changes only two special cases. If the input is +1, the "if" branch would be executed, and the current iteration breaks. If the input is -1, the same thing as M45 happens. For any other input, the program executes as it should. This means that M56 truly subsumes M45.
Figure~\ref{fig:testcode} shows the test case that kills M45, but not M56. The input value here is a single control sequence, which is neither -1 or +1, and therefore cannot kill M56. However, since it should have been handled by "else if" branch and M45 does not allow that, it kills M45. Hence, in Figure~\ref{fig:jterminal} (on the left side) , we can see that M56 dynamically subsumes M45.
Analysis such as this allows researchers to understand the relations between the mutants and reduce the effects of redundant mutants on their results.

\begin{table}
	\centering
	\caption{JTerminal Software Information}
	\label{table:cases}
	
	\adjustbox{max width=\linewidth}{
		\begin{tabular}{|c|c|c|c|c|c|c|c|c|c|}
			\hline \multirow{2}{*}{\textbf{Project}} & \multirow{2}{*}{\textbf{Ver.}} & \multicolumn{2}{c|}{\textbf{Size (LoC)}} & \multirow{2}{*}{\textbf{\#C}} & \multirow{2}{*}{\textbf{TS}} & \multirow{2}{*}{\textbf{SC}} & \multirow{2}{*}{\textbf{BC}} & \multirow{2}{*}{\textbf{MC}} & \multirow{2}{*}{\textbf{\#M}}\\
			\hhline{~~--~~~~} &  & \textbf{Prod.} & \textbf{Test} &  &  & & & & \\ 
			\hline
			\hline JTerminal  & 1.0.1 & 687 & 428  & 8 & 2 & 66\% & 56\% & 60.0\% & 160\\ 
			\hline
			\multicolumn{10}{c}{} \\
			\multicolumn{1}{l}{Acronyms:} & \multicolumn{9}{r}{Version (Ver.), Line of code (LoC), Production code (Prod.),} \\
			\multicolumn{10}{r}{Number of commits (\#C),  Team size (TS), Statement coverage (SC),}\\
			\multicolumn{10}{r}{Branch coverage (BC), Mutation coverage (MC), Number of Mutants (\#M)}
		\end{tabular}

	}
\end{table}

\section{Conclusion}
\label{S:Conclusion}
Many academic  studies in the field of software testing rely on mutation testing to use as their comparison criteria, and the existence of redundant mutants is a significant threat to their validity.  We developed a component for our mutation testing tool, LittleDarwin, to facilitate the detection of redundant mutants using dynamic  mutant subsumption analysis. 
We performed dynamic mutant subsumption analysis on a small, real-world project to demonstrate the capabilities of our tool. 
Using our tool, it is possible to detect and filter out redundant mutants, and help in increasing  the confidence in  results of experiments using  mutation testing as a comparison criteria.

\balance

\bibliographystyle{ACM-Reference-Format}
\bibliography{Master} 


\begin{thebibliography}{00}


\ifx \showCODEN    \undefined \def \showCODEN     #1{\unskip}     \fi
\ifx \showDOI      \undefined \def \showDOI       #1{{\tt DOI:}\penalty0{#1}\ }
  \fi
\ifx \showISBNx    \undefined \def \showISBNx     #1{\unskip}     \fi
\ifx \showISBNxiii \undefined \def \showISBNxiii  #1{\unskip}     \fi
\ifx \showISSN     \undefined \def \showISSN      #1{\unskip}     \fi
\ifx \showLCCN     \undefined \def \showLCCN      #1{\unskip}     \fi
\ifx \shownote     \undefined \def \shownote      #1{#1}          \fi
\ifx \showarticletitle \undefined \def \showarticletitle #1{#1}   \fi
\ifx \showURL      \undefined \def \showURL       {\relax}        \fi
\providecommand\bibfield[2]{#2}
\providecommand\bibinfo[2]{#2}
\providecommand\natexlab[1]{#1}
\providecommand\showeprint[2][]{arXiv:#2}

\bibitem[\protect\citeauthoryear{Ammann, Delamaro, and Offutt}{Ammann
  et~al\mbox{.}}{2014}]%
        {Ammann2014}
\bibfield{author}{\bibinfo{person}{P. Ammann}, \bibinfo{person}{M.~E.
  Delamaro}, {and} \bibinfo{person}{J. Offutt}.}
  \bibinfo{year}{2014}\natexlab{}.
\newblock \showarticletitle{Establishing Theoretical Minimal Sets of Mutants}.
  In \bibinfo{booktitle}{{\em 2014 IEEE Seventh International Conference on
  Software Testing, Verification and Validation}}. \bibinfo{pages}{21--30}.
\newblock
\showISSN{2159-4848}
\showDOI{%
\url{https://doi.org/10.1109/ICST.2014.13}}


\bibitem[\protect\citeauthoryear{Andrews, Briand, and Labiche}{Andrews
  et~al\mbox{.}}{2005}]%
        {Andrews2005}
\bibfield{author}{\bibinfo{person}{James~H. Andrews},
  \bibinfo{person}{Lionel~C. Briand}, {and} \bibinfo{person}{Yvan Labiche}.}
  \bibinfo{year}{2005}\natexlab{}.
\newblock \showarticletitle{Is Mutation an Appropriate Tool for Testing
  Experiments?}. In \bibinfo{booktitle}{{\em Proc. {ICSE} 2005 (27th
  international conference on software engineering)}} {\em
  (\bibinfo{series}{ICSE '05})}. \bibinfo{publisher}{ACM},
  \bibinfo{address}{New York, NY, USA}, \bibinfo{pages}{402--411}.
\newblock
\showISBNx{1-58113-963-2}
\showDOI{%
\url{https://doi.org/10.1145/1062455.1062530}}


\bibitem[\protect\citeauthoryear{Andrews, Briand, Labiche, and Namin}{Andrews
  et~al\mbox{.}}{2006}]%
        {Andrews2006}
\bibfield{author}{\bibinfo{person}{James~H. Andrews},
  \bibinfo{person}{Lionel~C. Briand}, \bibinfo{person}{Yvan Labiche}, {and}
  \bibinfo{person}{Akbar~Siami Namin}.} \bibinfo{year}{2006}\natexlab{}.
\newblock \showarticletitle{Using Mutation Analysis for Assessing and Comparing
  Testing Coverage Criteria}.
\newblock \bibinfo{journal}{{\em {IEEE} Transactions on Software
  Engineering\/}} \bibinfo{volume}{32}, \bibinfo{number}{8}
  (\bibinfo{date}{aug} \bibinfo{year}{2006}), \bibinfo{pages}{608--624}.
\newblock
\showISSN{0098-5589}
\showDOI{%
\url{https://doi.org/10.1109/tse.2006.83}}


\bibitem[\protect\citeauthoryear{Budd}{Budd}{1980}]%
        {Budd1980}
\bibfield{author}{\bibinfo{person}{Timothy~Alan Budd}.}
  \bibinfo{year}{1980}\natexlab{}.
\newblock {\em \bibinfo{title}{Mutation Analysis of Program Test Data}}.
\newblock \bibinfo{thesistype}{Ph.D. Dissertation}. \bibinfo{school}{Yale
  University}, \bibinfo{address}{New Haven, CT, USA}.
\newblock
\newblock
\shownote{AAI8025191.}


\bibitem[\protect\citeauthoryear{DeMillo, Lipton, and Sayward}{DeMillo
  et~al\mbox{.}}{1978}]%
        {DeMillo1978}
\bibfield{author}{\bibinfo{person}{Richard~A. DeMillo},
  \bibinfo{person}{Richard~J. Lipton}, {and} \bibinfo{person}{F.~G. Sayward}.}
  \bibinfo{year}{1978}\natexlab{}.
\newblock \showarticletitle{Hints on Test Data Selection: Help for the
  Practicing Programmer}.
\newblock \bibinfo{journal}{{\em Computer\/}} \bibinfo{volume}{11},
  \bibinfo{number}{4} (\bibinfo{date}{apr} \bibinfo{year}{1978}),
  \bibinfo{pages}{34--41}.
\newblock
\showISSN{0018-9162}
\showDOI{%
\url{https://doi.org/10.1109/C-M.1978.218136}}


\bibitem[\protect\citeauthoryear{Fawcett}{Fawcett}{2006}]%
        {Fawcett2006}
\bibfield{author}{\bibinfo{person}{Tom Fawcett}.}
  \bibinfo{year}{2006}\natexlab{}.
\newblock \showarticletitle{An introduction to {ROC} analysis}.
\newblock \bibinfo{journal}{{\em Pattern Recognition Letters\/}}
  \bibinfo{volume}{27}, \bibinfo{number}{8} (\bibinfo{date}{jun}
  \bibinfo{year}{2006}), \bibinfo{pages}{861--874}.
\newblock
\showISSN{0167-8655}
\showDOI{%
\url{https://doi.org/10.1016/j.patrec.2005.10.010}}
\newblock
\shownote{ROC Analysis in Pattern Recognition.}


\bibitem[\protect\citeauthoryear{Fraser and Arcuri}{Fraser and Arcuri}{2017}]%
        {SBST17_competition}
\bibfield{author}{\bibinfo{person}{Gordon Fraser} {and} \bibinfo{person}{Andrea
  Arcuri}.} \bibinfo{year}{2017}\natexlab{}.
\newblock \showarticletitle{EvoSuite at the SBST 2017 Tool Competition}. In
  \bibinfo{booktitle}{{\em 10th International Workshop on Search-Based Software
  Testing (SBST'17) at ICSE'17}}. \bibinfo{pages}{39--42}.
\newblock


\bibitem[\protect\citeauthoryear{Jia and Harman}{Jia and Harman}{2008}]%
        {Jia2008}
\bibfield{author}{\bibinfo{person}{Yue Jia} {and} \bibinfo{person}{Mark
  Harman}.} \bibinfo{year}{2008}\natexlab{}.
\newblock \showarticletitle{Constructing Subtle Faults Using Higher Order
  Mutation Testing}. In \bibinfo{booktitle}{{\em Proc. SCAM 2008 (Eighth {IEEE}
  International Working Conference on Source Code Analysis and Manipulation)}}.
  \bibinfo{publisher}{Institute of Electrical {\&} Electronics Engineers
  ({IEEE})}, \bibinfo{pages}{249--258}.
\newblock
\showDOI{%
\url{https://doi.org/10.1109/scam.2008.36}}


\bibitem[\protect\citeauthoryear{Jia and Harman}{Jia and Harman}{2011}]%
        {Jia2011}
\bibfield{author}{\bibinfo{person}{Yue Jia} {and} \bibinfo{person}{Mark
  Harman}.} \bibinfo{year}{2011}\natexlab{}.
\newblock \showarticletitle{An Analysis and Survey of the Development of
  Mutation Testing}.
\newblock \bibinfo{journal}{{\em {IEEE} Transactions on Software
  Engineering\/}} \bibinfo{volume}{37}, \bibinfo{number}{5}
  (\bibinfo{date}{sep} \bibinfo{year}{2011}), \bibinfo{pages}{649--678}.
\newblock
\showISSN{0098-5589}
\showDOI{%
\url{https://doi.org/10.1109/TSE.2010.62}}


\bibitem[\protect\citeauthoryear{Just, Jalali, Inozemtseva, Ernst, Holmes, and
  Fraser}{Just et~al\mbox{.}}{2014}]%
        {Just2014}
\bibfield{author}{\bibinfo{person}{Ren{\'e} Just}, \bibinfo{person}{Darioush
  Jalali}, \bibinfo{person}{Laura Inozemtseva}, \bibinfo{person}{Michael~D.
  Ernst}, \bibinfo{person}{Reid Holmes}, {and} \bibinfo{person}{Gordon
  Fraser}.} \bibinfo{year}{2014}\natexlab{}.
\newblock \showarticletitle{Are Mutants a Valid Substitute for Real Faults in
  Software Testing?}. In \bibinfo{booktitle}{{\em Proc. FSE 2014 ({Proceedings
  of the 22nd ACM SIGSOFT International Symposium on Foundations of Software
  Engineering})}} {\em (\bibinfo{series}{FSE 2014})}. \bibinfo{publisher}{ACM},
  \bibinfo{address}{New York, NY, USA}, \bibinfo{pages}{654--665}.
\newblock
\showISBNx{978-1-4503-3056-5}
\showDOI{%
\url{https://doi.org/10.1145/2635868.2635929}}


\bibitem[\protect\citeauthoryear{Kuhn}{Kuhn}{1999}]%
        {Kuhn1999}
\bibfield{author}{\bibinfo{person}{D.~Richard Kuhn}.}
  \bibinfo{year}{1999}\natexlab{}.
\newblock \showarticletitle{Fault Classes and Error Detection Capability of
  Specification-based Testing}.
\newblock \bibinfo{journal}{{\em ACM Trans. Softw. Eng. Methodol.\/}}
  \bibinfo{volume}{8}, \bibinfo{number}{4} (\bibinfo{date}{Oct.}
  \bibinfo{year}{1999}), \bibinfo{pages}{411--424}.
\newblock
\showISSN{1049-331X}
\showDOI{%
\url{https://doi.org/10.1145/322993.322996}}


\bibitem[\protect\citeauthoryear{Kurtz, Ammann, and Offutt}{Kurtz
  et~al\mbox{.}}{2015}]%
        {Kurtz2015}
\bibfield{author}{\bibinfo{person}{B. Kurtz}, \bibinfo{person}{P. Ammann},
  {and} \bibinfo{person}{J. Offutt}.} \bibinfo{year}{2015}\natexlab{}.
\newblock \showarticletitle{Static analysis of mutant subsumption}. In
  \bibinfo{booktitle}{{\em Software Testing, Verification and Validation
  Workshops (ICSTW), 2015 IEEE Eighth International Conference on}}.
  \bibinfo{pages}{1--10}.
\newblock
\showDOI{%
\url{https://doi.org/10.1109/ICSTW.2015.7107454}}


\bibitem[\protect\citeauthoryear{Mathur}{Mathur}{1991}]%
        {Mathur1991}
\bibfield{author}{\bibinfo{person}{Aditya~P. Mathur}.}
  \bibinfo{year}{1991}\natexlab{}.
\newblock \showarticletitle{Performance, effectiveness, and reliability issues
  in software testing}. In \bibinfo{booktitle}{{\em Proc. {COMPSAC 1991} (The
  Fifteenth Annual International Computer Software {\&} Applications
  Conference)}}. \bibinfo{publisher}{Institute of Electrical {\&} Electronics
  Engineers ({IEEE})}, \bibinfo{pages}{604--605}.
\newblock
\showDOI{%
\url{https://doi.org/10.1109/cmpsac.1991.170248}}


\bibitem[\protect\citeauthoryear{Offutt, Lee, Rothermel, Untch, and
  Zapf}{Offutt et~al\mbox{.}}{1996}]%
        {Offutt1996}
\bibfield{author}{\bibinfo{person}{A.~Jefferson Offutt}, \bibinfo{person}{Ammei
  Lee}, \bibinfo{person}{Gregg Rothermel}, \bibinfo{person}{Roland~H. Untch},
  {and} \bibinfo{person}{Christian Zapf}.} \bibinfo{year}{1996}\natexlab{}.
\newblock \showarticletitle{An Experimental Determination of Sufficient Mutant
  Operators}.
\newblock \bibinfo{journal}{{\em ACM Transactions on Software Engineering
  Methodology\/}} \bibinfo{volume}{5}, \bibinfo{number}{2}
  (\bibinfo{date}{April} \bibinfo{year}{1996}), \bibinfo{pages}{99--118}.
\newblock
\showISSN{1049-331X}
\showDOI{%
\url{https://doi.org/10.1145/227607.227610}}


\bibitem[\protect\citeauthoryear{Offutt and Pan}{Offutt and Pan}{1997}]%
        {Offutt1997}
\bibfield{author}{\bibinfo{person}{A.~Jefferson Offutt} {and}
  \bibinfo{person}{Jie Pan}.} \bibinfo{year}{1997}\natexlab{}.
\newblock \showarticletitle{Automatically detecting equivalent mutants and
  infeasible paths}.
\newblock \bibinfo{journal}{{\em Software Testing, Verification and
  Reliability\/}} \bibinfo{volume}{7}, \bibinfo{number}{3} (\bibinfo{date}{sep}
  \bibinfo{year}{1997}), \bibinfo{pages}{165--192}.
\newblock
\showDOI{%
\url{https://doi.org/10.1002/(sici)1099-1689(199709)7:3<165::aid-stvr143>3.0.co;2-u}}


\bibitem[\protect\citeauthoryear{Offutt, Rothermel, and Zapf}{Offutt
  et~al\mbox{.}}{1993}]%
        {Offutt1993}
\bibfield{author}{\bibinfo{person}{A.~Jefferson Offutt}, \bibinfo{person}{Gregg
  Rothermel}, {and} \bibinfo{person}{Christian Zapf}.}
  \bibinfo{year}{1993}\natexlab{}.
\newblock \showarticletitle{An Experimental Evaluation of Selective Mutation}.
  In \bibinfo{booktitle}{{\em Proc. ICSE 1993 (15th international conference on
  Software engineering)}} {\em (\bibinfo{series}{ICSE '93})}.
  \bibinfo{publisher}{IEEE Computer Society Press}, \bibinfo{address}{Los
  Alamitos, CA, USA}, \bibinfo{pages}{100--107}.
\newblock
\showISBNx{0-89791-588-7}
\showURL{%
\url{http://dl.acm.org/citation.cfm?id=257572.257597}}


\bibitem[\protect\citeauthoryear{Offutt and Untch}{Offutt and Untch}{2001}]%
        {Offutt2001}
\bibfield{author}{\bibinfo{person}{A.~Jefferson Offutt} {and}
  \bibinfo{person}{Roland~H. Untch}.} \bibinfo{year}{2001}\natexlab{}.
\newblock \showarticletitle{Mutation 2000: Uniting the Orthogonal}.
\newblock In \bibinfo{booktitle}{{\em Mutation Testing for the New Century}},
  \bibfield{editor}{\bibinfo{person}{W.Eric Wong}} (Ed.). \bibinfo{series}{The
  Springer International Series on Advances in Database Systems},
  Vol.~\bibinfo{volume}{24}. \bibinfo{publisher}{Springer US},
  \bibinfo{pages}{34--44}.
\newblock
\showISBNx{978-1-4419-4888-5}
\showDOI{%
\url{https://doi.org/10.1007/978-1-4757-5939-6_7}}


\bibitem[\protect\citeauthoryear{Papadakis, Henard, Harman, Jia, and
  Le~Traon}{Papadakis et~al\mbox{.}}{2016}]%
        {Papadakis2016}
\bibfield{author}{\bibinfo{person}{Mike Papadakis},
  \bibinfo{person}{Christopher Henard}, \bibinfo{person}{Mark Harman},
  \bibinfo{person}{Yue Jia}, {and} \bibinfo{person}{Yves Le~Traon}.}
  \bibinfo{year}{2016}\natexlab{}.
\newblock \showarticletitle{Threats to the Validity of Mutation-based Test
  Assessment}. In \bibinfo{booktitle}{{\em Proceedings of the 25th
  International Symposium on Software Testing and Analysis}} {\em
  (\bibinfo{series}{ISSTA 2016})}. \bibinfo{publisher}{ACM},
  \bibinfo{address}{New York, NY, USA}, \bibinfo{pages}{354--365}.
\newblock
\showISBNx{978-1-4503-4390-9}
\showDOI{%
\url{https://doi.org/10.1145/2931037.2931040}}


\bibitem[\protect\citeauthoryear{Parsai}{Parsai}{2015}]%
        {Parsai2015T}
\bibfield{author}{\bibinfo{person}{Ali Parsai}.}
  \bibinfo{year}{2015}\natexlab{}.
\newblock {\em \bibinfo{title}{Mutation Analysis: An Industrial Experiment}}.
\newblock \bibinfo{thesistype}{Master's\ thesis}. \bibinfo{school}{University
  of Antwerp}.
\newblock


\bibitem[\protect\citeauthoryear{Parsai, Murgia, and Demeyer}{Parsai
  et~al\mbox{.}}{2016a}]%
        {Parsai2016}
\bibfield{author}{\bibinfo{person}{Ali Parsai}, \bibinfo{person}{Alessandro
  Murgia}, {and} \bibinfo{person}{Serge Demeyer}.}
  \bibinfo{year}{2016}\natexlab{a}.
\newblock \showarticletitle{Evaluating Random Mutant Selection at Class-level
  in Projects with Non-adequate Test Suites}. In \bibinfo{booktitle}{{\em Proc.
  {EASE} 2016 (20th International Conference on Evaluation and Assessment in
  Software Engineering)}} {\em (\bibinfo{series}{EASE '16})}.
  \bibinfo{publisher}{ACM}, \bibinfo{address}{New York, NY, USA}, Article
  \bibinfo{articleno}{11}, \bibinfo{numpages}{10}~pages.
\newblock
\showISBNx{978-1-4503-3691-8}
\showDOI{%
\url{https://doi.org/10.1145/2915970.2915992}}


\bibitem[\protect\citeauthoryear{Parsai, Murgia, and Demeyer}{Parsai
  et~al\mbox{.}}{2016b}]%
        {Parsai2016M}
\bibfield{author}{\bibinfo{person}{Ali Parsai}, \bibinfo{person}{Alessandro
  Murgia}, {and} \bibinfo{person}{Serge Demeyer}.}
  \bibinfo{year}{2016}\natexlab{b}.
\newblock \showarticletitle{A Model to Estimate First-Order Mutation Coverage
  from Higher-Order Mutation Coverage}. In \bibinfo{booktitle}{{\em Proc. {QRS}
  2016 ({IEEE} International Conference on Software Quality, Reliability and
  Security)}}. \bibinfo{publisher}{Institute of Electrical and Electronics
  Engineers ({IEEE})}, \bibinfo{pages}{365--373}.
\newblock
\showDOI{%
\url{https://doi.org/10.1109/QRS.2016.48}}


\bibitem[\protect\citeauthoryear{Parsai, Murgia, and Demeyer}{Parsai
  et~al\mbox{.}}{2017}]%
        {Parsai2017}
\bibfield{author}{\bibinfo{person}{Ali Parsai}, \bibinfo{person}{Alessandro
  Murgia}, {and} \bibinfo{person}{Serge Demeyer}.}
  \bibinfo{year}{2017}\natexlab{}.
\newblock \showarticletitle{LittleDarwin: a Feature-Rich and Extensible
  Mutation Testing Framework for Large and Complex Java Systems}. In
  \bibinfo{booktitle}{{\em Proc. FSEN 2017 (Fundamentals of Software
  Engineering)}}.
\newblock


\bibitem[\protect\citeauthoryear{Wong and Mathur}{Wong and Mathur}{1995}]%
        {Wong1995}
\bibfield{author}{\bibinfo{person}{W.~Eric Wong} {and}
  \bibinfo{person}{Aditya~P. Mathur}.} \bibinfo{year}{1995}\natexlab{}.
\newblock \showarticletitle{Reducing the cost of mutation testing: An empirical
  study}.
\newblock \bibinfo{journal}{{\em Journal of Systems and Software\/}}
  \bibinfo{volume}{31}, \bibinfo{number}{3} (\bibinfo{year}{1995}),
  \bibinfo{pages}{185--196}.
\newblock
\showISSN{0164-1212}
\showDOI{%
\url{https://doi.org/10.1016/0164-1212(94)00098-0}}


\bibitem[\protect\citeauthoryear{Zhang, Gligoric, Marinov, and Khurshid}{Zhang
  et~al\mbox{.}}{2013}]%
        {Zhang2013}
\bibfield{author}{\bibinfo{person}{Lingming Zhang}, \bibinfo{person}{Milos
  Gligoric}, \bibinfo{person}{Darko Marinov}, {and} \bibinfo{person}{Sarfraz
  Khurshid}.} \bibinfo{year}{2013}\natexlab{}.
\newblock \showarticletitle{Operator-based and random mutant selection: Better
  together}. In \bibinfo{booktitle}{{\em Proc. ASE 2013 ({28th IEEE/ACM
  International Conference on Automated Software Engineering})}}.
  \bibinfo{publisher}{Institute of Electrical {\&} Electronics Engineers
  ({IEEE})}, \bibinfo{pages}{92--102}.
\newblock
\showDOI{%
\url{https://doi.org/10.1109/ASE.2013.6693070}}


\bibitem[\protect\citeauthoryear{Zhang, Hou, Hu, Xie, and Mei}{Zhang
  et~al\mbox{.}}{2010}]%
        {Zhang2010}
\bibfield{author}{\bibinfo{person}{Lu Zhang}, \bibinfo{person}{Shan-Shan Hou},
  \bibinfo{person}{Jun-Jue Hu}, \bibinfo{person}{Tao Xie}, {and}
  \bibinfo{person}{Hong Mei}.} \bibinfo{year}{2010}\natexlab{}.
\newblock \showarticletitle{Is Operator-based Mutant Selection Superior to
  Random Mutant Selection?}. In \bibinfo{booktitle}{{\em Proc. ICSE 2010 vol. 1
  ({Proceedings of the 32nd ACM/IEEE International Conference on Software
  Engineering - Volume 1})}} {\em (\bibinfo{series}{ICSE '10})}.
  \bibinfo{publisher}{ACM}, \bibinfo{address}{New York, NY, USA},
  \bibinfo{pages}{435--444}.
\newblock
\showISBNx{978-1-60558-719-6}
\showDOI{%
\url{https://doi.org/10.1145/1806799.1806863}}


\end{thebibliography}

\end{document}